\documentstyle[psfig]{article}
\begin{document}
\renewcommand{\theequation}{\thesection.\arabic{equation}}
\def\bq{\begin{equation}}
\def\eq{\end{equation}}
\def\bqa{\begin{eqnarray}}
\def\eqa{\end{eqnarray}}

\vspace{2cm}

\title{The Continuum Version of $\phi^4_{1+1}$-theory in
Light-Front Quantization}
\author{Pierre GRANG\'E$^{(1)}$, Peter ULLRICH$^{(2)}$ 
and Ernst WERNER$^{(2)}$\\
(1)Laboratoire de Physique Math\'ematique et The\'orique\\
Universit\'e Montpellier II \\
F-34095, Montpellier, Cedex 05 - FRANCE \\
(2)Instit\"ut f\"ur Theoretische Physik\\
D-93040, Universit\"at Regensburg - GERMANY}
\date{}
\maketitle

\vspace{2.5cm}

\begin{abstract}
A genuine continuum  treatment of the massive $\phi^4_{1+1}$-theory in
light-cone quantization is proposed. Fields are treated as operator
valued distributions thereby leading to a mathematically well defined
handling of ultraviolet and light cone induced infrared divergences
and of their renormalization. Although non-perturbative the continuum
light cone approach is no more complex than usual perturbation theory
in lowest order. Relative to discretized light cone quantization,
the critical coupling increases by $30\%$ to a value $r = 1.5$. 
 Conventional perturbation theory at the corresponding order 
yields $r_1=1$,
 whereas the
 RG improved fourth order result is $r_4 = 1.8 \pm 0.05$.
\end{abstract}

\vfill
\hrule\vskip10pt
\noindent{\small PM 97/18, June 1997\hfill
}

\section{Introduction}
The discretized light front quantization (DLCQ) [1] has played an important
role in clarifying infrared aspects of the theory which are decisive for
the appearance of the vacuum sector field, the LC-counterpart of
the nontrivial ground state of ET-quantization [2-7]. The popularity of 
DLCQ resides in the easy and conceptually simple treatment of the
necessary infrared regularisation. However it has never been demonstrated
that the limit where the periodicity length L goes to infinity is
identical to the genuine continuum theory where momentum space
discretization is avoided from the start. The reason lies in the infrared
behaviour of the continuum theory which has not yet been understood. Our
aim is to clarify this issue on the basis of a mathematically well
defined procedure.

As an example we treat explicitely the $\phi^4_{1+1}$-theory in
the continuum and compare its results for the phase transition to the
DLCQ case. It turns out that with the same type of physical
approximations the characteristics of the phase transition are the same
in both cases whereas the critical coupling strength and the dependence
of the order parameter on the coupling strength are substantially
different.

 In connection with phase transitions there is a vital interest to
dispose of a continuum version of the theory, if one is interested in
the study of critical phenomena in the framework of effective theories.
This is the point of view of statistical theories of fields in which
cutoffs are introduced in order to define a momentum or mass scale below
which the effective theory is valid. Critical points of phase
transitions are determined from zeros of the $\beta$-function. To do
this requires the complete knowledge of the cutoff dependence of the
critical mass which can be given only by the continuum theory.

In Section 2 we make use of  the concept of field operator valued
distributions to have a  mathematically
well defined  Fock expansion. 
This can be done in a chart-independent manner expressing the field as a
surface integral over a manifold. Comparing the expressions for the
Minkowski 
and light-front cases one sees that the regularization properties of
the test functions are automatically transferred from the first to the
second case ; thus it is ensured that, if the field is regular in the 
Minkowski
case, it is also regular in the LC-case - it is the same field  
expressed by different surface integrals ! 
Actually what is called IR-divergence in the unregularized LC-field
expansion is an UV-divergence in the LC-energy; it is only the special
choice of coordinates which makes it look like an IR-divergence.
Therefore there is no extra
infrared (IR) singularity in the LC-case which would have to be treated
separately : the ultraviolet (UV) behaviour of the field on the Minkowski
manifold dictates the UV \underline{and} IR behaviour on the LC
manifold. There is absolutely no freedom in the LC-case beyond the
choice of test functions relevant in the Minkowski case. Moreover, due
 to
general properties known from functional analysis the independence of
physical results from the special form of the test functions is ensured.
In Section 3 we use the Haag expansion of field operators to define the
decomposition into the particle sector field $\varphi$ and the vacuum
sector field $\Omega.$
 In Section 4 we discuss the equation of motion
(EM) for $\varphi$ and the constraint for $\Omega$ which are coupled
equations.
 Finally in Section 5 we discuss the so called mean field
solution of these equations and the results for the phase transition.
In Section 6 we rewrite the condition for the phase transition in the
language of an effective theory and compare the results with the
literature.
In Appendix A we collect a number of results from the theory of
distributions to substantiate the discussions in Section 2.

\section{Fundamentals. Definitions and Conventions}

The physical system under consideration is defined by the Lagrangian

\bq {\cal L} = {1 \over 2} (\partial_{\mu} \phi)(\partial^{\mu} \phi) -
 {m^2 \over 2} \phi^2(x) - {\lambda \over 4!} \phi^4(x).
\ \ \ m^2 > 0 \ \ \ ; \ \ \ \lambda > 0. \label{aa} \eq

From (\ref{aa}) follows the EM

\bq \partial_{\mu} \partial^{\mu} \phi + m^2 \phi + {\lambda \over 3 !}
\phi^3 = 0. \label{ac} \eq

The free field EM defines the free field $\varphi_0(x)$ which can be
expanded in terms of free field creation and annihilation operators.

 We
start with the Minkowski case (ET quantization). The field
$\varphi_0(x)$ is defined in the sense of distributions by the functional 

\bq \varphi_0\{u\} = (\varphi_0(x), u(x)) = \int d^4x \varphi_0(x) u(x)
\label{ad} \eq
where $u(x)$ is an element of the test function space in coordinate spaces.
Plugging in the Fock expansion of $\varphi_0(x)$ yields

\bq (\varphi_0(x), u(x)) = {1 \over (2\pi)^3} \int {d^4 x d^3p  \over 2 \omega
(\vec{p})}. 
[a(\vec{p})  e^{-i<x,p>_M}  + a^+(\vec{p}) e^{+i<x,p>_M}] .
f(\omega(\vec{p}), \vec{p}) \label{ae} \eq
where $<x,p>_M$ is the scalar product of $x$ and $p$ in the Minkowski metric,\\
$\omega(\vec{p}) = \sqrt{m^2 +\vec{p}^2}$ is the on-shell energy and $f(p_0,
\vec{p})$ is a test function in momentum space which is required to fall off
sufficiently fast as a function of the arguments
 $p_0,p_1,p_2,p_3$ as any one of the
$p_i's$ goes to $\infty$. 

The minimal conditions for the attenuation factor $f(p)$ are

$$\int {d^3p \over 2 \omega(p)} |f(\omega(\vec{p}), \tilde p)||<s|a^+
(\vec{p})|s'>| < \infty $$
and
$$  \int {d^3p \over 2 \omega(p)}|f(\omega(\tilde p), \vec{p})| |<r|a 
(\vec{p})|
r'>| < \infty. $$
Here $(|s>, |s'>)$ and $(|r>, |r'>)$ are arbitrary pairs of states yielding 
nonvanishing matrix
elements for $a^+$ and $a$ respectively (see App. eq. (A.15)).
This condition is necessary, if one wants to garantee that the Fourier
integral in  (\ref{ae}) is finite. $\varphi_0(x)$ is then defined as

\bq \varphi_0(x) = {1 \over (2\pi)^3} \int {d^3p \over 2 \omega(\vec{p})}  
[a (\vec{p}) e^{-i<x,p>_M} + a^+ (\vec{p}) e^{i<x,p>_M}] f(p_0, \vec{p})
\label{af} \eq
which can also be written as a surface integral over the manifold defined by
$p^2_0 - \vec{p}^2 - m^2 = 0$ (see App. A). The positive and negative
frequency parts in (\ref{ae}) and (\ref{af}) are a consequence of the
necessity to introduce two charts - corresponding to $p_0 = \pm \sqrt{p^2 +
m^2}$ - if one wants to cover the manifold.

In the light-cone case the corresponding expression becomes

\bq \varphi_0(x) = {1 \over (2 \pi)^3} \int d^3 \tilde p {\theta(p^+) \over
2p^+} [\tilde a (\vec {\tilde p}) e^{-i<\tilde x, \tilde p>_{LC}} + \tilde a^+ 
(\vec {\tilde p}) e^{i<\tilde x, \tilde p>_{LC}}] \tilde f(p_0 (\tilde p), 
\vec {p}(\tilde p)). \label{ag} \eq

Here $\tilde{x}$ and $\tilde{p}$ designate the light-cone variables

$$ \tilde{x}^0 : = x^+ = {1 \over \sqrt2}(x^0 + x^3)$$

$$ \tilde{x}^3 : = x^- = {1 \over \sqrt{2}}(x^0 - x^3)$$

\bq \tilde{x}^2 : = x^i \ \ \ , \ \ \ i = 1,2 .\label{ah} \eq

$$ \tilde{p}^3: = p^+ = {1 \over \sqrt{2}} (p^0 + p^3)$$

$$\tilde{p}^0: = p^- = {1 \over \sqrt{2}} (p^0 - p^3)$$

\bq \tilde{p}^i = p^i \ \ \ ; \ \ \ i = 1,2 \label{ai} \eq
 $<\tilde x, \tilde p>_{LC}$ is the scalar  product within the LC-metric. The
creation and annihilation operators in (\ref{af}) and (\ref{ag}) are related
through

\bq \tilde a (\vec{\tilde p}) = a (\vec{v}(\vec{\tilde p})) \ \ ; \ \
 \tilde a^+ (\vec{\tilde p}) = a^+ (\vec{v}(\vec{\tilde p})) \ \ ;\ \ 
\forall \vec{\tilde p}\ \ |p^+ > 0\label{aj} \eq
where $\vec{v}(\vec{\tilde p})$ is the three vector defined by
 
\bq \vec{v}(\vec{\tilde p}) = (\vec{p}_{\perp}, {1 \over\sqrt{2}}(p^+ -
{(p^2_{\perp} + m^2 \over 2p^+}))\label{ak} \eq
Finally $\tilde{f}(p_0 (\vec{p}), \vec{p}(\tilde p))$ is the transformed test
function.

Explicitely we have

\bq \tilde{f} = f[{1 \over \sqrt{2}} (p^+ + {p^2_{\perp} + m^2 \over 2p^+}), (
p^1_{\perp}, p^2_{\perp}, {1 \over \sqrt{2}} (p^+ - {p^2_{\perp} + m^2
 \over 2p^+})] \label{al} \eq
from where it is clear that there is no infrared singularity in (\ref{ag}), if
there is none in (\ref{af}), since the singular behaviour of ${1 \over p^+}$ is
completely damped out by the behaviour of the test function for $p^+ \to 0$.
This is also clear from the fact that the two integrals in Eqs. (\ref{af}) and 
(\ref{ag}) are equal (see App. A). Therefore, if $\varphi_0(x)$ is a
bounded operator in the ET case it is also guaranteed to be bounded in the 
light
cone case. Whereas the integral in the Minkowski case is composed of
contributions from two charts, the final expression  (\ref{ag}) in the LC-case
goes only over one chart (the one with $p^+ >0$). Originally there were also
two charts, corresponding to $p^+ < 0$ and $p^+ > 0$, but the integral over
the former one turns out to be equal to the integral over 
$p^+ > 0$ hence merging in a single expression
 ; this is  due to the
different topologies in the Minkowski and the LC-case : in the LC-case the
sign of $p^-$ is the same as the sign of $p^+$, whereas in the Minkowski case
the sign of the energy is not correlated with signs of momentum components ;
instead there is a sign ambiguity.

\setcounter{equation}{0}

\section{Decomposition of fields into particle sector and vacuum 
sector fields}

In the DLCQ-case  the total field $\phi(x)$ can be
naturally decomposed into the particle sector $\varphi(x)$  - to be
constructed from polynomials in $\tilde{a}^+(k^+)$ and
 $\tilde{a}(k^+)$ with total
momentum $K^+ > 0$ - and the vacuum sector $\Omega$ - to be constructed
from polynomials with total momemtum  $K^+ = 0$ [4]. In the continuum case
this decomposition can be achieved with the help of the Haag expansion
[8].

As in the DLCQ-case we decompose $\phi(x) = \varphi(x) + \Omega$ where
for fixed LC time we have :

$$ \varphi(x) = \varphi_0(x) + \int dy^-_1 dy^-_2 g_2(x^- - y^-_1, x^-
-y^-_2) : \varphi_0(y_1) 
\varphi_0(y_2) :$$
\bq  + \int dy^-_1 dy^-_2 dy^-_3 g_3(x^- - y^-_1,x^- - y^-_2 , x^--y_3)
 : \varphi_0(y_1)
\varphi_0(y_2) \varphi_0(y_3): + ... \label{ca} \eq

All fields are taken at a fixed time, e.g. $x^+ = 0$ ; the argument $y_i$ means
therefore $y_i = (y^-_i, x^+)$.

Due to the properties of $\varphi_0(x)$ the support of $\varphi$ in
Fourier space is determined by the support of the test functions in
$\varphi_0(x)$. The coefficient functions $g_2,g_3,,....$ - or 
rather their Fourier transforms -  have to be
determined from the equation of motion and the constraints.

In order to obtain the vacuum sector field $\Omega$ - which by
definition is $x^-$ independent - we perform an additional integration over
$x^-$ and add a constant c-number part $\phi_0$  :

$$\Omega : = \phi_0 + {1 \over V} \int dx^- dy^-_1 dy^-_2 g_2(x^- -y^-_1,x^- - 
y^-_2) : 
\varphi_0(y_1) \varphi_0(y_2)
: $$
\bq + {1 \over V} \int dx^- dy^-_1 dy^-_2 g_3(x^- -y^-_1, x^- -y^-_2, x^- - 
y^-_3) 
: \varphi_0(y_1)
\varphi_0(y_2) \varphi_0(y_3) : +... , \label{cb} \eq

V being the integration volume.

Apparently the operator valued part of $\Omega$ is nonlocal. Due to the
fact that $\varphi_0$ is defined as an operator valued distribution the
integrations in (\ref{ca}) and (\ref{cb}) are well-defined.

Substituting the expansion (\ref{ag})  into the definition (\ref{cb})
one obtains after a lengthy but completely standard calculation the
Fourier expansion of $\Omega$ :

\bq \Omega = \phi_0 + \int^{\infty}_0 {dk^+ \over 4\pi k^+} f^2
(k^+,\hat{k}^-(k^+)) C(k^+) \tilde{a}^+(k^+)\tilde{a}(k^+) + ... \label{cc}
\eq
where the coefficient $C(k^+)$ is given by

$$ C(k^+) = {2 \over V} \int \int g_2(x^--y^-_1, x^- - y^-_2) \cos [{k^+
\over 2} (y_2-y_1)] dy_1 dy_2$$

The higher terms of the expansion are not reproduced here because they
will not be considered in this paper. Apparently one has as in DLCQ :

$$ <0 |\Omega|0> \ \  = \ \ <0 |\phi|0> = \phi_0$$
and

\bq \Omega |q^+_1, q^+_2 ,... q^+_N> = \lambda(q^+_1 ... q^+_N) |q^+_1,
q^+_2,... q^+_N> \label{cd} \eq
the eigenvalues $\lambda(q^+_1 ... q^+_N)$ being given by 

\bq \lambda(q^+_1 ... q^+_N) = \phi_0 + \sum^N_{i=1} {f^2(q^+_i,
\hat{q}^-(q^+_i)) \over 4 \pi q^+_i} C(q^+_i). \label{ce} \eq

This shows that within the bilinear approximation $\Omega$ acts 
like a momemtum dependent mass term.

\setcounter{equation}{0}

\section{Determination of $\phi_0$ and $C(k^+)$}

The field $\phi(x) = \varphi(x) + \Omega$ satisfies
 the LC-form of the equation of
motion

\bq 2 \partial_+ \partial_-(\varphi(x) + \Omega) + m^2 (\varphi(x) +
 \Omega) + {\lambda \over 3!} (\varphi(x) + \Omega)^3 = 0. \label{da} \eq

We first define an operator $P$ which projects an operator $I\!\!F(x)$
onto the vacuum sector according to

\bq P * I\!\!F(x) : = {1 \over V} \int^{+ \infty}_{- \infty}
I\!\!F(x) dx. \label{db} \eq

Acting with $P$ on (\ref{da}) yields the constraint (the
derivative term vanishes)

\bq \theta_3 : = m^2 \Omega + {\lambda \over 3!} \Omega^3 
 + {\lambda \over 3!} {1 \over V} \int^{+ \infty}_{- \infty}
 [\varphi^3(x)\Omega + \Omega + \varphi^2(x) + \varphi(x)
 \Omega \varphi(x)]dx = 0.
\label{dc} \eq

Projection with the complementary operator $ Q : ={\bf 1} - P$ yields the
equation of motion for $\varphi(x)$ :

\bq 2 \partial_+\partial_- \varphi(x) + m^2\varphi(x) + {\lambda \over
3!} Q * (\varphi(x) + \Omega)^3 = 0, \label{dd} \eq
(\ref{dc}) and (\ref{dd}) are coupled operator valued equations which
are solved by taking matrix elements between Hilbert space states.
Technically this is very similar to the DLCQ case [4] :

One replaces $\varphi \to \varphi_0$ in (\ref{dc}) and (\ref{cb}) and
calculates the matrix elements \\ $<0 | \theta_3 | 0>$ and $<k^+ |
\theta_3 | k^+>$. 

The results are
\bq < 0 | \theta_3 | 0> = \mu^2 \phi_0 + {\lambda \over 3!} \phi_0^3 +
{\lambda \over 24\pi} \int^{\infty}_0 dk^+ {C(k^+) \hat{f}^4(k^+) \over k^+} = 
0.
\label{de} \eq
and
$$<k^+ |\theta_3 | k^+> = {\lambda \over 6} C^3(k^+) \hat{f}^6(k^+) +
{\lambda \over 2} \phi_0 C^2(k^+) \hat{f}^4(k^+) $$

$$+ [\mu^2 \hat{f}^2(k^+) + {\lambda \over 2} \phi^2_0 \hat{f}^2(k^+) + 
{\lambda \over 4\pi k^+}\hat{f}^4(k^+)] C(k^+)$$

\bq + {\lambda \phi_0 \over 4\pi k^+} \hat{f}^2(k^+) = 0 \ \ ; 
\ \ \forall \ \
k^+ \label{df} \eq

Here we use the notation $f(k^+, \hat{k}^-(k^+)) : = \hat{f}(k^+).$ $\mu^2$ is
defined by

\bq \mu^2 = m^2 + {\lambda \over 8 \pi} \int^{\infty}_0 {dk^+ \over k^+} 
\hat{f}^2
(k^+) \label{dg} \eq 
which  is nothing but the tadpole renormalization of the mass.In order to
 keep things as close as possible to the DLCQ case we use
dimensionless momenta which we measure in units of  $\mu$.

Apparently (\ref{de}) is an equation for $\phi_0$ as a functional of
$C(k^+)$ ; in turn (\ref{df}) determines $C(k^+)$ as a function of
$\phi_0$ in the form of a cubic equation. Whereas the "exact" solution
of eqs. (\ref{de}) and (\ref{df}) has to found numerically, important
qualitative 
features of the solution can be discussed  analytically.

In the region where $k^+$ is very small but $\hat{f}(k^+) \approx 1$
the solution of eq. (\ref{df}) is simply

\bq C(k^+) = - {\lambda \phi_0 \over 4 \pi k^+} \displaystyle{f^2(k^+)
\over {\lambda \over 4 \pi k^+} . f^4(k^+)} \approx - {\phi_0 \over
f^2(k^+)} \label{dh} \eq

i.e. the order parameter $\phi_0$ determines the infrared behaviour of
the vacuum vector part of the field.

 Using (\ref{dh}) in the small $k^+$ region in (\ref{de}) yields an
integrand which behaves in the IR as $\displaystyle{- \lambda \phi_0 \over 24 
\pi}
\hat{f}^2(k^+)/k^+$ which has, up to numerical factors, the same behavior
as  the tadpole contribution in eq. (\ref{dg}). Therefore, any
divergence in (\ref{de}) arising from sending cutoffs present in
$\hat{f}(k^+)$ to infinity can be cured by an appropriate mass
counterterm. It is important to note that this result is independent
of the particular form chosen for the test function. In the UV region
the integral in (\ref{de}) causes no problem whatsoever, since for $k^+
\to \infty \ \ \  C(k^+) \sim {1 \over k^+}$ which yields an integrand $\sim
1/k^{+^2}.$

Nevertheless, in order to be able to evaluate the integrals one has to
make a choice for $\hat{f}(k^+) = f(k^+, \hat{k}^-(k^+))$ ; without the
on-shell condition $\hat{k} = \hat{k}^- (k^+)$ the test functions
depend on the two variables $k^+$ and $k^-$. $f$ can be chosen to have
compact support and to be unlimitedly   differentiable. In our case the
support is the interior of a circle of radius  $\Lambda$ - which
later on will be identified with a cutoff - . A possible form is [9]

$$\rho(k^+,k^-) = \exp({1 \over \Lambda^2}) \exp [-
 {1 \over \Lambda^2 - (k^{+2}
 + k^{-2})}] \ \ ; \ \ k^{+2} + k^{-2} \leq \Lambda^2 $$

$$ = 0 \ \ \ \ \ \ \ k^{+2} + k^{-2} > \Lambda^2$$

From this function one can construct another one which has the
property that it equals 1 inside the 2-sphere of radius
$\Lambda - \epsilon$ and falls to zero within the interval
$[\Lambda-\epsilon, \Lambda] ; \epsilon$ can be chosen as small as one
likes without affecting the $C^{\infty}$-character of the function
[9].\\
In order to make the comparison with the DLCQ-case we use from now on
the convention $\hat{k}^- = m^2/k^+$ instead of $\hat{k}^- =  {m^2
\over 2k^+}$ (see section 2). This means the use of the factor $1 \over
2$ in eqs (\ref{ah}) and (\ref{ai}) instead of $1/\sqrt{2}$.

Taking into account the on-shell condition we arrive at the situation
depicted in fig. 1.

\vspace{-5cm}

\hspace*{2.5cm}
\psfig{file=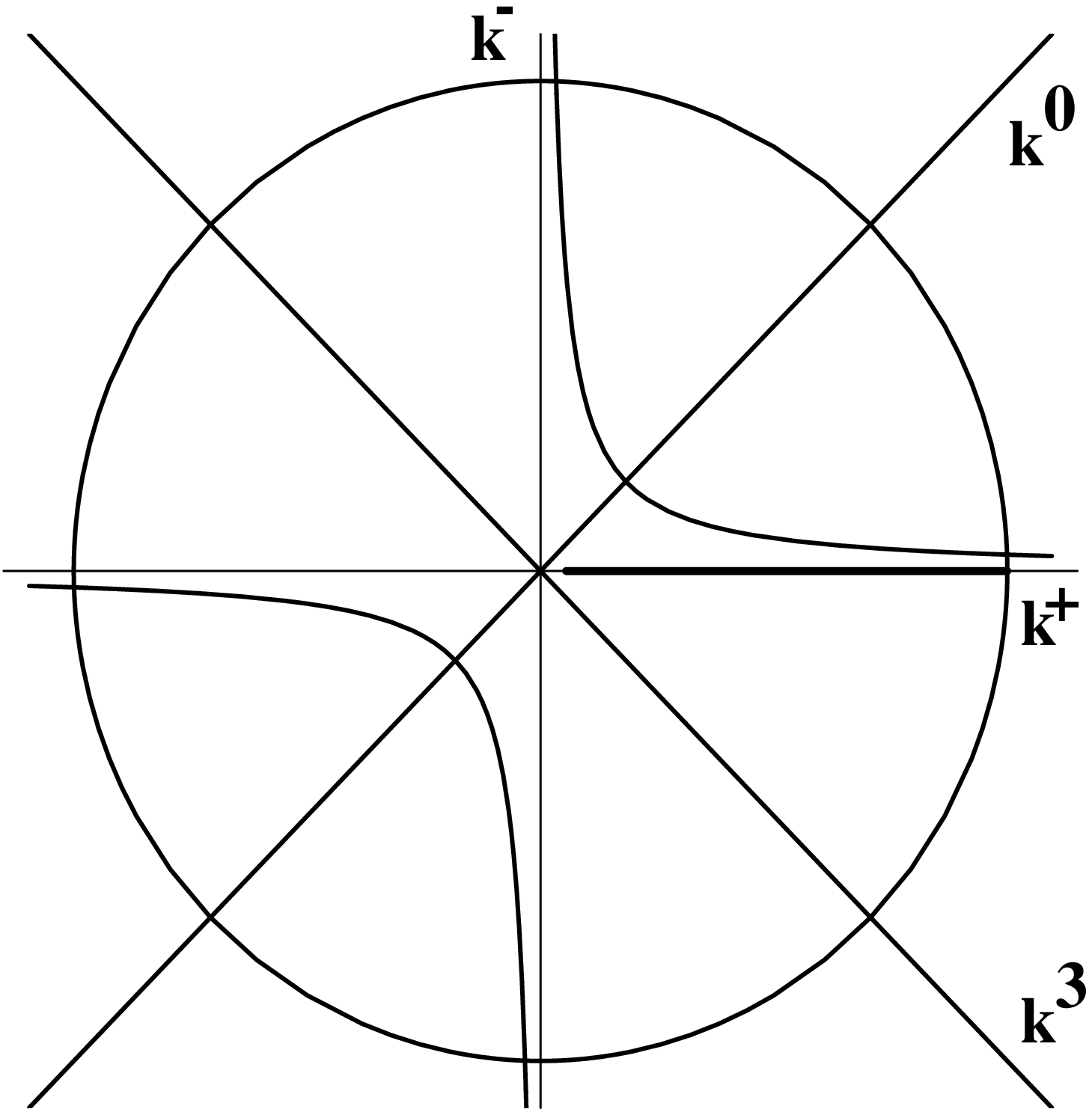,height=10cm}

\vspace{-1cm}\hspace*{-0.2cm}{FIG. 1. The area inside the circle is
 the support of the test
function $f(k^+,k^-)$. f equals unity inside a circle of radius
$\Lambda-\epsilon$. The fall-off to zero takes place in the
interval$[\Lambda-\epsilon,\Lambda]$. The hyperbola represents the
on-shell condition $\hat{k}^- = m^2/k^+$. Its intersections with the
circle determine the IR and UV cut-offs for the variable $k^+$. The
resulting support for the function $\hat{f}(k^+)$ is indicated by the
thick line. Only the right half is physically realized due to the
kinematical condition $k^+>0$}\\  

The final result for $\hat{f}(k^+)$ is shown in fig. 2.

\vspace{-2.0cm}\hspace*{2.5cm}\psfig{file=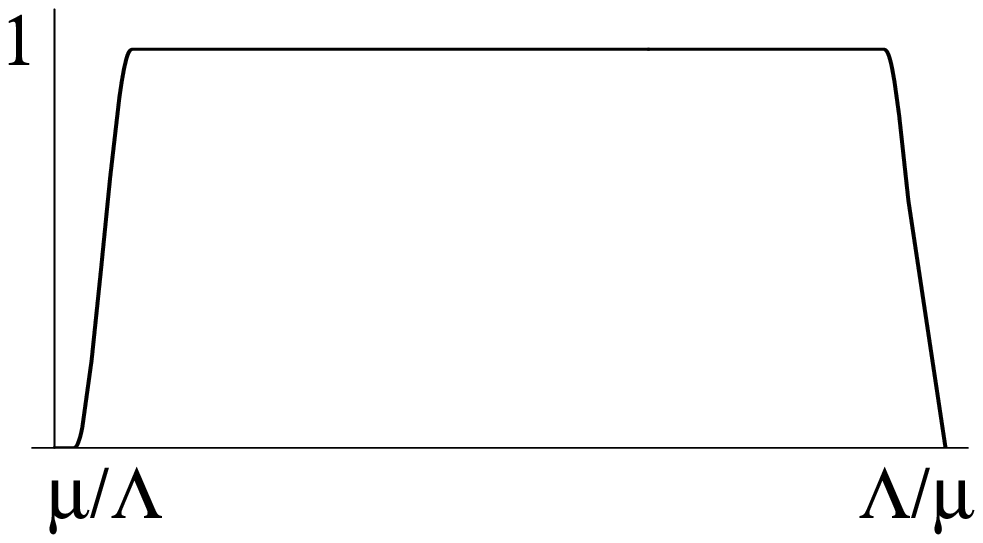,height=9cm,width=7cm}

\vspace{-1.5cm}\hspace*{2.5cm}{FIG. 2. Generic form of the test
function$\hat{f}(k^+)$}\\

In the limit where $\epsilon$ is arbitrarely small $\hat{f}$ acts like
a cutoff at $1/\Lambda$ and $\Lambda$.
 
Using this cutoff form does not influence the physical results, since any
other test function having the same support would yield the same
results. Of course, instead of the divergent integral in (\ref{di}) we
would get something else but the counter-term of eq. (\ref{dh}) would change
correspondingly.We have tested and verified this statement by doing numerical
integrations with small, but finite values of $\epsilon.$

Using from now on the cutoff form for $\hat{f}$ and the dimensionless
coupling $g = \lambda/4\pi \mu^2$ the integral to be renormalized
becomes

\bq I(\Lambda) = - {g \over 6} \int^{\Lambda/ \mu}_{\mu/ \Lambda} dk^+ {C(k^+)
\over k^+} \label{di} \eq
which is $UV$ finite but IR divergent. Using the IR limit of $C(k^+)$
given in eq(\ref{dh}) $I(\Lambda)$ becomes

\bq I(\Lambda) = - {g \phi_0 \over 6} \log ({\Lambda \over \mu})
 - {\phi_0 \over 6(\Lambda/\mu)}
 + 0({\mu^2 \over \Lambda^2}) ...\ \ . \label{dj} \eq
Given that

\bq < 0 |\varphi^2_0| 0> = {1 \over 8 \pi} \int^{\Lambda /\mu}
_{\mu/ \Lambda}
{dk^+ \over k^+} = {1 \over 4\pi} \log (\Lambda /\mu) \label{dk} \eq
the divergence in (\ref{di}) can be compensated by the substraction of
a mass-type counter-term $2 \pi g \phi_0 \varphi^2_0/3.$

\setcounter{equation}{0}
 
\section{Critical coupling and nature of the phase transition}

In the vicinity of the phase transition where $ \phi_0 \ll 1$ one can
linearize  eqs. (\ref{de}) and (\ref{df}) yielding

\bqa C(k^+) &=& - {\lambda \phi_0 \over 4\pi k^+} {1 \over \mu^2 + {\lambda
\over 4\pi k^+}} \nonumber \\
            &=& - g \phi_0 {1 \over (g+k^+)}.
\label{ea} \eqa

The phase transition being determined by the vanishing of the mass term the
critical coupling $g_c$ is determined by the condition

\bq 1 = {g_c^2 \over 6} \int^{\Lambda/\mu}_{\mu/\Lambda} {dk^+
 \over k^+(k^++g_c)} -
{2\pi \over 3} g_c <0 |\varphi_0^2 | 0> \label{eb} \eq

which follows from (\ref{de}) after division by $\mu^2$ and substraction of
the mass counterterm.

The integral in (\ref{eb}) can be evaluated analytically and yields with
(\ref{dk}) :

$${g_c \over 6} \log (g_c) = 1 $$
which has the solution $g_c = 4.19...$

The corresponding value for DLCQ [4] is $g_{c_{DLCQ}} = 3.18$, i.e. there is a 
30\%
deviation between the two cases. On the other hand there is no change in the
nature of the phase transition (which is of second order) and of the critical
exponents.

\setcounter{equation}{0}
\section{Comparison to theories of critical behaviour}

In order to compare our value for the critical coupling $g_c$ to
results obtained earlier in ET-quantization we have to rewrite the
constraint (\ref{de}). The most complete study of the critical
behaviour of $\phi^4_{1+1}$-theory has been performed by Parisi [10] in the
scenario of a theory of critical phenomena. In this context the field
theory is interpreted as an effective theory with a cutoff $\Lambda$
which defines the scale of validity of the theory. In the spirit of
such a theory one has to keep in the constraints (\ref{de}) the
dependence on $\Lambda$ and consider this equation as a prescription
for the calculation of the critical mass $M(\tilde{g},\Lambda).$ From
this quantity one obtains the $\beta$-function

\bq \beta(\tilde{g}) = M {\partial M \over \partial
\tilde{g}}|_{\lambda, \Lambda}. \label{fa} \eq
Here the definition of the coupling $\tilde{g}$ differs from our $g$ -
all momenta and masses are measured in units of $\Lambda$, distances
are replaced by the dimensionless quantity $x \Lambda$ - . $g$ and 
$\tilde{g}$ are related by

\bq g = {\lambda \over 4 \pi \mu^2} = {\lambda \over 4 \pi \Lambda^2}
{\Lambda^2 \over \mu^2} : = \tilde{g} {\Lambda^2 \over \mu^2}
\label{fb} \eq

We consider the constraint $\theta_3$ as an equation for $M^2$ :

\bq M^2 = \mu^2 + {\lambda \over 24 \pi} \int^{\Lambda / \mu}_{{\mu/
 \Lambda}} {C(k^+)\over k^+} dk^+ \label{fc} \eq

Satisfying the constraint $\theta_3 = 0$ amounts to $M^2(\Lambda,
\tilde{g}) = 0$ which in turn means $\beta(\tilde{g},\Lambda) = 0$ i.e.
the condition which defines the phase transition via the fixed point of
the $\beta$-function.

We define the critical mass $\mu_c$ by
\bq 0 = \mu^2_c + {\lambda \over 24 \pi} \int^{\Lambda / \mu_c}_{\mu_c
/ \Lambda} {C(k^+) \over k^+} dk^+ = \mu^2_c + {\lambda \over 24
\pi} [ \log {g_c + k^+ \over k^+}]|^{\Lambda / \mu_c}
_{\mu_c / \Lambda} \label{fd} \eq

Using $g_c = \tilde{g}_c {\Lambda^2 \over \mu^2_c}$ and the new variable
$y(\tilde{g}_c) = \log({\Lambda \over
\mu_c})^2$, we obtain in the limit of large $\Lambda$ 

\bq y \displaystyle e^{y}  = {6 \over \tilde{g}_c} \label{fe} \eq
The numerical solution of eq. (\ref{fe}) is shown in fig.3 as a
function of $\tilde{g}_c.$
\vspace*{-1.5cm}

\hspace*{1.5cm}\psfig{file=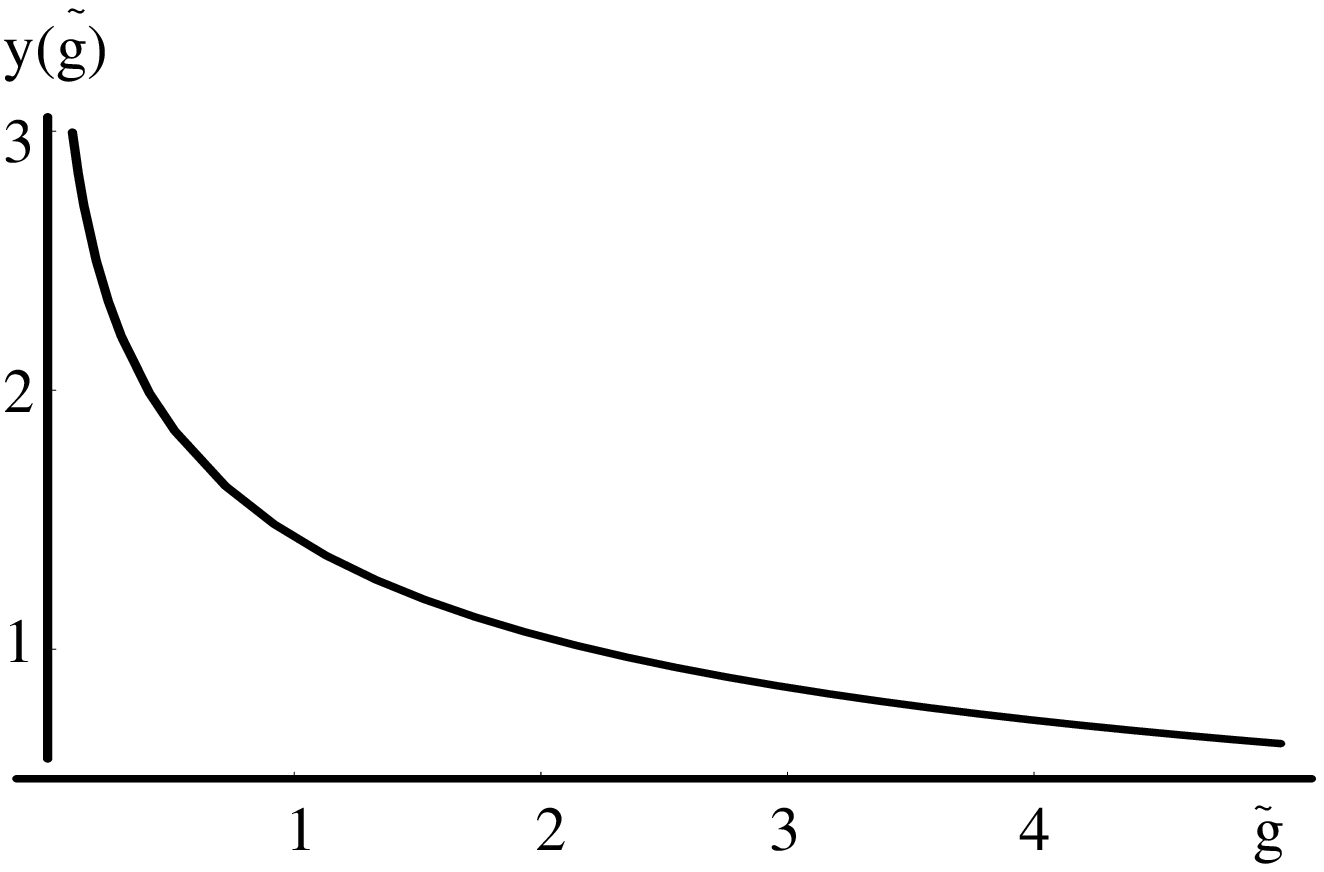,height=10.2cm,width=9.2cm}

\vspace{-2cm}\hspace*{2.5cm}{FIG. 3. The function $y(g)$ solution of
eq. (\ref{fe}).} \\

Knowing $y(\tilde{g}_c)$ we can now relate $\tilde{g}_c$ and $g_c$ by

$$ \tilde{g}_c = g_c ({\mu_c \over \Lambda})^2 = g_c  \displaystyle e^{-y
(\tilde{g}_c)}$$
or
\bq \tilde{g}_c \displaystyle e^{y(\tilde{g}_c)} = g_c \label{ff} \eq

In fig. 4 the left hand side of eq. (\ref{ff}) is shown together with 
the straight line corresponding to $g_c = 4.19$ .\\

\vspace{-2.0cm}\hspace*{1.5cm}\psfig{file=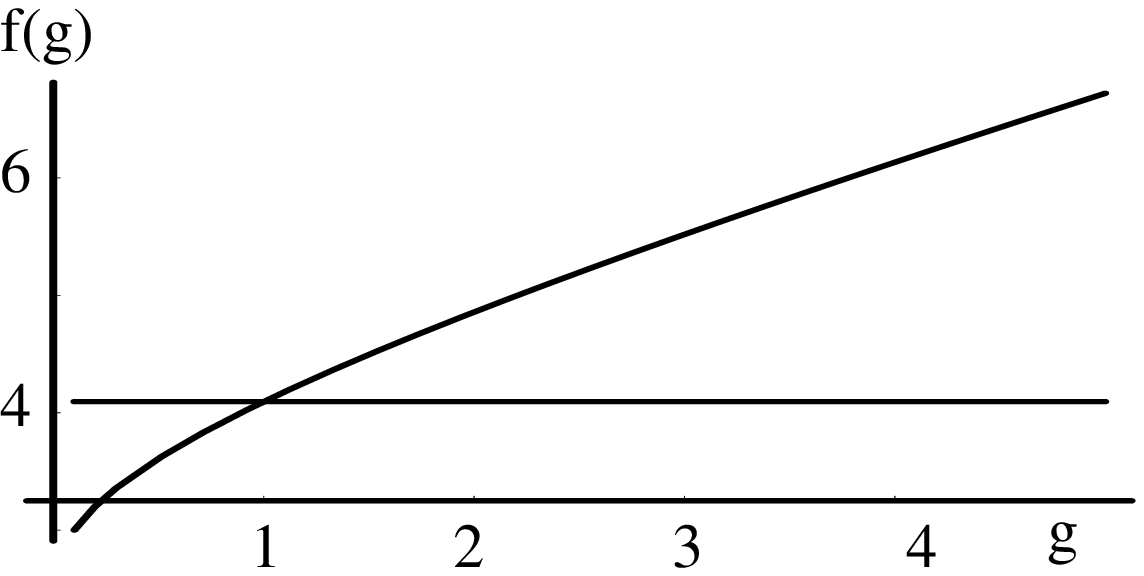,height=10.2cm,width=9.2cm}

\vspace{-2cm}\hspace*{3.5cm}{FIG. 4. The solution of
eq. (\ref{ff}).} \\

The numerical value for $\tilde{g}_c$ is 1.

Parisi [10] uses in his calculation still another coupling, called $r$,
defined by $r = {3 \lambda \over 8\pi \Lambda^2} = {3 \over 2} \tilde{g}$. It
is normalized in such way that the critical coupling at the
order of one-loop is $r_1 = 1.0$.
 He pushed his calculations up to four loops (with a Borel-improvement
of the convergence of the asymptotic series) and obtained the value
for $r_4$ reproduced in tab.1.

\vspace{1cm}

\hspace*{2.5cm}\psfig{file=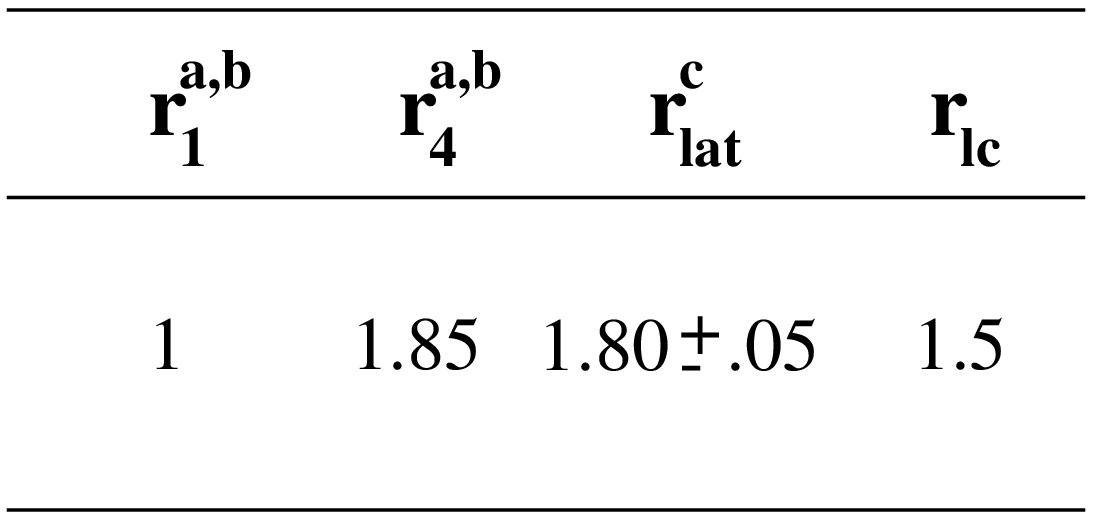,height=3cm,width=8cm}

\hspace*{3.5cm}{a) Ref.[10]}\hspace*{0.5cm}{b) Ref.[11]}\hspace*{0.5cm}{c) 
Ref.[12]}

\vspace{1cm}\hspace*{1.0cm}{TAB. 1. Critical couplings by different methods. 
$r_{lc}$
 is the value from\\ }
\hspace*{3cm}{the present continuous light-cone calculation}

\vspace{1cm}

As far as the solution of the equation of motion is concerned our
result corresponds to the 1-loop result of Parisi et al. (tadpole
correction of the mass). On the other hand it is nonperturbative as far
as the solution of the constraint is concerned. This is reflected by
the considerable improvement relative to $r_1 = 1$ which brings us
with $r_{lc} = 1.5$ already rather close to the four-loop result $r_4 =
1.85$ and to the lattice result $r_{lat} = 1.80 \pm 0.05$.

\section{Conclusions}

We have shown for scalar fields that the continuum quantum field theory
quantized on the light-cone does not suffer from divergence problems
beyond those present in conventional quantization, if the field operators
are treated properly as operator valued distributions. The treatment is
quite generic and should be rather easily generalizable to other types of
fields. Apart from giving substantially different results for the critical
coupling of the $\phi^4_{1+1}$-theory as compared to the DLCQ, the
continuum version has the advantage that it can be rewritten as an
effective theory for critical phenomena. This is important for a detailed
comparison with the literature because the most elaborate studies in
conventional quantization have been performed in this domain. Our results
compare very favorably with the best values of RG-improved fourth order
perturbation theory and of lattice calculations which are reached up to $20
\%$. Given the calculational simplicity of our approach - which on the
technical level corresponds to first order perturbation theory - this is
an
encouraging success. It is attributed to the existence of an operator
valued vacuum sector field which has to be added to the usual particle
sector field and which is the LC-signature of nonperturbative physics.

\vspace{0.5cm}
\noindent{\bf Acknowledgement :}{\it This work has been supported by the NATO
grant $n^o$ CRG920472.}

\vfill\eject

\setcounter{equation}{0}

\section*{Appendix A}

\vspace{1cm}

In this appendix we review some concepts from the theory of distributions
which are  at the basis of the results of Section 2.

\subsection*{A.1 - Pullback of distributions}

Here we can give only a very short version. For more details the reader
can consult e.g. reference [13].

We consider 2 open subsets $U$ and $V$ of $R^n : U, V \subset {\cal R}^n$ and
a  $C^{\infty}$-diffeomorphism $\kappa$ between them :

$$\kappa : U \to V .$$
Given a distribution

$$(\varphi, f) = \int_V \varphi(x) f_V(x) dx \eqno{(A.1)}$$
with $f_V(x)$ a test function of compact support on $V$ the distribution
$\varphi$ can be "pulled back" to $U$ with the help of the pullback mapping
$\kappa^* = \kappa \circ \varphi = \varphi(\kappa(x))$ (see fig. )

$$ (\kappa^*[\varphi],f) = \int_U \varphi(\kappa(x)) f_U(x) dx
 \eqno{(A.2)}$$\\

\hspace*{3.5cm}\psfig{file=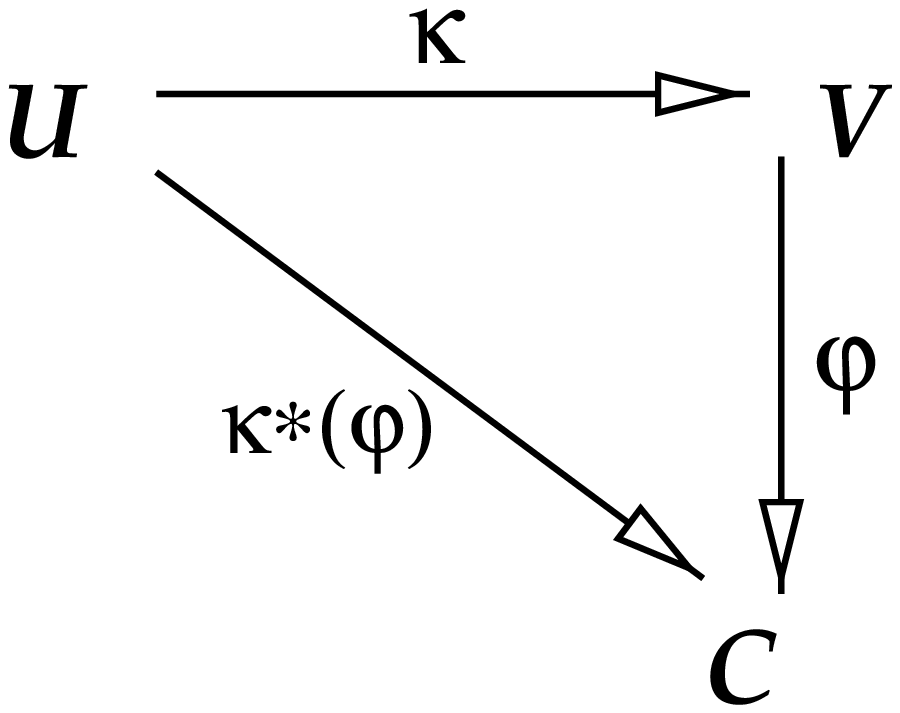,height=4cm,width=5cm}

$$\kappa^* : C(V) \to C(U) \ \ \ ; \ \ \ \kappa^*(\varphi): = \varphi \circ
\kappa $$

Introducing the inverse mapping

$$\tau: = \kappa^{-1} : V \to U$$
and making the coordinate transformation $\xi : = \kappa(x)$ (A.2) goes
over into

$$ (\kappa^*[\varphi],f) = \int_V \varphi(\xi) f(\tau (\xi)) | \det
D(\tau(\xi))| d\xi  \eqno{(A.3)}$$
where $D(\tau(\xi))$ is the Jacobian of the coordinate transformation.
(A.2)
assigns  to each distribution $\varphi$ on $V$ a "pulled back" distribution
$\kappa^*[\varphi]$ on $U$ and (A.3) is the prescription for its
evaluation in terms of $\varphi$. The applications which one has in mind in
connection with (A.2) and (A.3) is the frequent case where one works
with distributions which depend on functions of the integration variables,
e.g. $\delta(p^2-m^2)$.

In order to treat such a case in the pullback framework we take :\\
 $U \subset
{\cal R}^n, n >      1 \ \ ; \ \ V \subset {\cal R}^1$ is obtained from $U$ 
through
the mapping $Q : U \to V$ with a $C^{\infty}$-function which we call now $Q$
instead of $\kappa$. Moreover we introduce the (n-1)-dimensional sub-manifold
$\Sigma$ defined by $Q(x) = 0$ ;

$$\Sigma  : = \{x \in U | Q(x) = 0\} $$
and a Dirac $\delta^1$-distribution $\delta_0$ in ${\cal R}^1$.

The following  theorem can be proved [13] :\\
If
$ \bigtriangledown Q(x) \neq 0 \ \ \ \ \forall x \in U$
then the pullback of $\delta^1$ exists and is defined by  

$$ \delta \circ Q = {1 \over |\bigtriangledown Q|} \delta_{\Sigma},  
\eqno{(A.4)}$$
where  $\delta_{\Sigma}$ is a distribution defined by

$$(\delta_{\Sigma}, f) = \int_{\Sigma} f dS \ \ \ \forall f \in U,
 \eqno{(A.5)}$$
here $dS$ is the euclidian surface measure on $\Sigma$ ; this yields
 the distribution $\delta o Q$ as a surface integral :

$$ (\delta \circ Q,f) = \int_{\Sigma} {f \over |\bigtriangledown Q|}dS 
\eqno{(A.6)}$$
Specializing $Q$ to $Q(p) = p^2 - m^2$, $p$ being the energy-momemtum
4-vector and using on $\Sigma$ the charts

$$ \Omega_{\pm} (\pm \omega(\vec p), \vec p) $$ 
the distribution (A.6) can be rewritten as (in Minkowsky space) :

$$ (\delta \circ Q, f) = \int {d^3 p \over 2 \omega (\vec{p})}
[f(\omega(\vec{p}), \vec{p}) + f(-\omega(\vec{p}), \vec{p})]
\eqno{(A.7)}$$
where $\omega (\vec{p}) = + \sqrt{p^2+m^2}.$

We introduce the two tempered distributions

$$  \delta_{\pm}(p^2-m^2) = 1(\vec{p}) \theta (\pm p^0) \delta(p^2-m^2),
 \eqno{(A.8)}$$
where $1(\vec{p})$ is the unit distribution defined by  
$$(1,f) = \int f(\vec{p}) d^3 p, $$
and $\delta_{\pm}$ obey 

$$ (\delta_{\pm},f) = \int {d^3 p \over 2 \omega (\vec{p})} f(\pm
\omega (\vec{p}), \vec{p}). \eqno{(A.9)}$$

Comparison with (A.7) shows that

$$(\delta \circ Q)(p) = \delta_+ (p^2 - m^2) + \delta_-(p^2 - m^2).$$

Going back to (A.6) in the form 

$$(\delta_+ (p^2 - m^2) + \delta_-(p^2 - m^2), f(p)) = (\delta \circ Q,f(p))
= \int_{\Sigma^+} {f \over |\nabla Q|} ds +\int_{\Sigma^-}  
{f \over |\nabla Q|} ds$$

we see that $\delta_+$ and $\delta_-$ can be written as surface
integrals.

$$(\delta_{\pm}(p^2 - m^2), f(p)) = \int_{\Sigma^{\pm}}
 {f \over |\nabla Q|} ds. 
 \eqno{(A.10)}$$

Here the two integrals $\int_{\Sigma^{\pm}} ds$ are over the two surfaces
defined by the two signs of $p_0$ in $p_o = \pm \sqrt{\vec{p}^2 + m^2}$
with charts $\Omega^{\pm}$ (A.5). It is important to note that the
integrals in (A.10) are independent of the special choice of the charts
$\Omega^{\pm}$ which one makes on the surfaces $\Sigma^{\pm}$. 

\subsection*{A.2 - Solutions of the KG-equation in Minkowski space}

The tempered
distribution $\chi(p) \delta(p^2-m^2)$ defined by

$$\chi(p) \delta (p^2 - m^2) = \chi_+(p) \delta_+(p^2 - m^2) + \chi_-(p)
\delta_-(p^2 - m^2)  \eqno{(A.11)}$$
satisfies

$$ (\chi(p) \delta (p^2 - m^2), f(p)) = \int_{\Sigma} {\chi f \over 
|\bigtriangledown
Q|} ds  \eqno{(A.12)}$$

and solves the KG equation in momentum space :
$$ (p^2 - m^2) v(p) = 0$$
with
$$v(p) = v_1 (\vec{p}) \theta(p^0) \delta(p^2 - m^2) + v_2(\vec{p}) 
\theta(-p^0)  \delta(p^2 - m^2).$$
From the distribution $\chi(p) \delta(p^2 - m^2)$ one obtains the solution
of the coordinate space KG-equation as a distribution $\phi(x)$ defined by
(Minkowsky metric)
$$ \phi(x) := 2\pi {\cal F}_{M_x} [\chi(p) \delta(p^2 - m^2)]  \eqno{(A.13)}$$
where $\cal{F}_M$ symbolises the Fourier-transform with Minkowsky metric, or
more explicitely

$$ (\phi(x),f(x))  =  2 \pi (\chi(p) \delta(p^2-m^2), {\cal F}_M(f) (p))$$

$$ = 2 \pi \int_{\Sigma} {\chi ({\cal F}_M f) \over |\bigtriangledown Q|} ds $$

$$  = 2 \pi \int {d^3 p \over 2 \omega (\vec{p})} [\chi (\omega (\vec{p}),
\vec{p}). ({\cal F}_M f) (\omega (\vec{p}), \vec{p}) 
 +\chi (-\omega (\vec{p}),
\vec{p}) ({\cal F}_M f)(-\omega (\vec{p}), \vec{p})]. \eqno{(A.14)}$$

 In order to guarantee the existence of the two last integrals, one has to
impose the condition on $\chi$ :

$$ \int {|\chi| \over |\bigtriangledown Q|} ds < \infty\ \ \mbox{and} \ \ 
\int {d^3 p \over
2 \omega (\vec{p})} |\chi (\pm \omega (\vec{p}), \vec{p})| < \infty. 
\eqno{(A.15)}$$

 As it stands this is valid for classical fields.
After quantization - where the $\chi's$ are replaced  by creation and
annihilation operators  -  (A.15) becomes a condition for matrix elements of
these operators.

With (A.15) the functions

$$ H_{\pm} (x,p) = {1 \over 2 \omega (\vec{p})} \chi(\pm \omega 
(\vec{p}),
\vec{p}) e^{-i <x, \hat{p}_{\pm}>_M} f(x)$$
$$\hat{p}_{\pm} = (\pm \omega (\vec{p}),\vec{p})$$
are integrable ; consequently one can inject the Fourier representation of
${\cal F}_M f$ into (A.14) to obtain the regular distribution

$$ (\phi(x), f(x)) = {1 \over (2 \pi)^3} \int d^4x f(x) \int{dp^3 \over 
2\omega (\vec{p})} \ \
[\chi (\hat{p}_+) e^{-i <x, \hat{p}_+>_M} + \chi(\hat{p}_-)
e^{-i <x, \hat{p}_->_M}]  \eqno{(A.16)}$$
From (A.16) one identifies the field $\phi(x)$ as

$$ \phi = {1 \over (2 \pi)^3} \int{dp^3 \over 2\omega (\vec{p})}
[\chi (\hat{p}_+) e^{-i <x, \hat{p}_+>_M} + \chi(\hat{p}_-)
e^{-i <x, \hat{p}_->_M}]  \eqno{(A.17)}$$
the chart independent form of (A.17) is

$$ \phi = {1 \over (2 \pi)^3} \int_{\Sigma}{\chi (p) \over |\bigtriangledown 
Q(p)|} 
e^{-i <x, \hat{p}>_M}  ds.  \eqno{(A.18)}$$

\subsection*{A.3 - Solutions of the KG-equation on the LC}

We go from the coordinates $x,p$ to the corresponding LC coordinates
$\tilde{x}, \tilde{p}$ via the mapping

$$ \tilde{x} = \kappa \circ x \ \ ; \ \ \tilde{p} = \kappa \circ p$$
which leads to the LC-KG equation

$$ (\Box_{LC} + m^2) \tilde{\phi}(\tilde{x}) = 0$$
with
$$ \Box_{LC} = 2 \partial_+ \partial_- - \partial^2_\perp \ \ ;$$
Introducing the quadratic form

$$ \tilde{Q} (\tilde{p}) := \tilde{p}^2 - m^2  \eqno{(A.19)}$$
one can define the distribution $\delta(\tilde{p}^2 - m^2)$ as the pullback
of the $\delta$-distribution under the mapping (A.19) : 

$$ \delta(\tilde{p}^2 - m^2) := \delta \circ \tilde{Q} $$
$$ (\delta(\tilde{p}^2 - m^2), f(\tilde{p})) = \int_{\tilde{\Sigma}}
{f(\tilde{p}) \over |\bigtriangledown \tilde{Q}|} d\tilde{s}  \eqno{(A.20)}$$
where $\bigtriangledown \tilde{Q}(\tilde{p}) = 2 (p^- , - \vec{p}_{\perp},
p^+) \neq 0$ on the mass shell. The manifold $\tilde{\Sigma}$ is a sum of
two disconnected parts $\tilde{\Sigma}^+, p^+ > 0$ and 
$\tilde{\Sigma}^-, p^- < 0.$

Choosing on $\tilde{\Sigma}^{\pm}$ the charts
$$(\tilde{\Omega}^+ : (\tilde{\omega} (\vec{\tilde{p}}), \vec{\tilde{p}}) \ \
; \ \  p^+ > 0)$$
and
  $$(\tilde{\Omega}^- : (\tilde{\omega} (\vec{\tilde{p}}), \vec{\tilde{p}}) \ \
; \ \  p^+ < 0)$$
$$ \tilde{\omega} (\vec{\tilde{p}}) = {p^2_{\perp} + m^2 \over 2p^+}$$
one obtains :
$$(\delta (\tilde{p}^2-m^2), f(\tilde{p})) = \int_{\Sigma^+} 
{f(\tilde{p}) \over |\bigtriangledown \tilde{Q}|} d\tilde{s}
+ \int_{\Sigma^-} {f(\tilde{p}) \over |\bigtriangledown \tilde{Q}|} 
d \tilde{s} $$
$$ = \int d^3 \tilde{p} {\theta (p^+) \over 2 |p^+|} f
 (\tilde{\Omega}^+
(\vec{\tilde{p}})) + \int d^3  \tilde{p}{\theta (-p^+) \over 2 |p^+|} 
f(\tilde{\Omega}^- (\vec{\tilde{p}}))  \eqno{(A.21)}$$

The two integrals in (A.13) (with $\chi = 1)$ and (A.20) can be
shown to be equal. More generally it can be shown that for $\chi \in
{\cal L}^1(\Sigma)$ and $\tilde{\chi} = \chi \ \ \circ \ \ u^{-1} \in {\cal
 L}^1
(\tilde{\Sigma})$ one has the identity 

$$ \int_{\Sigma^{\pm}} {\chi(p) \over |\bigtriangledown Q|}ds = 
\int_{\tilde{\Sigma}^{\pm}} {\tilde{\chi}(\tilde{p}) \over 
|\bigtriangledown \tilde{Q}|}d \tilde{s}.  \eqno{(A.22)}$$
Using this identity one arrives immediately at the result (see eq.
(A.17)) :

$$ \phi(x) = {1 \over (2 \pi)^3} \int_{\Sigma} {\chi (p) \over 
|\bigtriangledown  Q(p)|}e^{-i <x, \hat{p}>_M} ds = {1 \over
 (2 \pi)^3}
\int_{\tilde{\Sigma}} {\tilde{\chi} (\tilde{p}) \over 
|\bigtriangledown \tilde{Q} (\tilde{p})|}e^{-i <\hat{x},
\hat{p}>_L} d \tilde{s}.  \eqno{(A.23)}$$

Introducing the same charts as in (A.21) one obtains

$$ \tilde{\phi} (\tilde{x}) = {1 \over (2 \pi)^3} \int d^3 p {1 \over
2|p^+|} \tilde{\chi}(\tilde{\Omega} (\tilde{p}))e^{-i
<\tilde{\Omega}
(\tilde{p}), \tilde{x}>_L}  \eqno{(A.24)}$$
where

$$\tilde{\chi}(\tilde{\Omega}(\vec{\tilde{p}})) = \chi [{1 \over \sqrt{2}}
(p^+ + {p^2_{\perp} + m^2 \over 2p^+}), \vec{p}_{\perp}, {1 \over
\sqrt{2}} (p^+ - {p^2_{\perp} + m^2 \over 2p^+})]. $$

The transition to the quantized field is made in (A.17) by the
substitution

$$\chi(\hat{p}_+) \to a(\vec{p})$$
$$\chi(- \hat{p}_-) \to a^+(\vec{p})$$
yielding
$$\hat{\phi}(x) = {1 \over (2 \pi)^3} \int {d^3p \over 2 \omega (\vec{p})}
[a(p)e^{-i <x, \hat{p}_->_M} + a^+(\vec{p})e^{i<x,\hat{p}_+>_M}].
$$

In the LC-case the two contributions from $\tilde{p}^+ > 0$ and 
$\tilde{p}^+ < 0$ in (A.23) can be recast into a single one by first
changing the integration variable in the second term $\vec{\tilde{p}}
\to - \vec{\tilde{p}}$ and then making the substitution

$$ \tilde{\chi}(\tilde{\Omega} (\vec{\tilde{p}})) \to \tilde{a}
(\vec{\tilde{p}}) \ \ ; \ \ \tilde{p}^+ > 0$$

$$ \tilde{\chi}(\tilde{\Omega}  (-  \vec{\tilde{p}})) \to \tilde{a}^+
(\vec{\tilde{p}}) \ \ ; \ \ \tilde{p}^+ > 0$$
yielding

$$ 
\hat{\tilde{\phi}}(\tilde{\chi}) = {1 \over (2 \pi)^3} \int
{d^3\tilde{p}} {\theta(\tilde{p}^+)\over 2\tilde{p}^+} [\tilde{a}
(\vec{\tilde{p}})e^{-i <\tilde{\Omega}
(\tilde{p}), \tilde{x}>_L} + \tilde{a}^+ (\vec{\tilde{p}})
e^{+i<\tilde{\Omega}(\tilde{p}), \tilde{x}>_L}]. 
 \eqno{(A.25)}$$

For completeness we add a remark : In the literature [14]
operator valued distributions have been introduced   on the level of
the Fock-space operators $a(p), a^+(p)$ by defining distributions

$$ (a,f) = {1 \over (2 \pi)^3} \int {d^3p \over 2 \omega (\vec{p})}
a(\vec{p}) f(\omega(\vec{p}), \vec{p}) $$

$$ (a^+,f) = {1 \over (2 \pi)^3} \int {d^3p \over 2 \omega (\vec{p})}
a^+(\vec{p})f(\omega(\vec{p}), \vec{p})  \eqno{(A.26)}$$

For the LC-case this reads

$$ (\tilde{a}, \tilde{f}) = {1 \over (2 \pi)^3} \int {d^3 \tilde{p} \over
\tilde{p}^+} \tilde{a} (\vec{\tilde{p}}) \tilde{f} (\omega(\vec{p}), \vec{p})$$

$$ (\tilde{a}^+, \tilde{f}) = {1 \over (2 \pi)^3}
 \int {d^3 \tilde{p} \over\tilde{p}^+}  \tilde{a}^+ (\vec{\tilde{p}})
 \tilde{f} (\omega(\vec{p}), \vec{p}  \eqno{(A.27)}$$

Decomposing the field $\varphi_0(x)$ into positive and negative
frequency parts $\varphi_0(x) = \varphi^+_0(x) + \varphi^-_0(x)$ we see
that
 
$$\varphi_0^+(0) = (a,f) = (\tilde{a}, \tilde{f})$$
$$\varphi_0^-(0) = (a^+,f) = (\tilde{a}^+, \tilde{f})$$
This shows again that the field $\varphi_0(x)$ is well defined in the
sense of operator valued distributions.

\vfill\eject

\noindent{\bf REFERENCES}

\begin{description}
\item[1.] H.C. Pauli, S.J. Brodsky, Phys.Rev. {\bf D32}, 1993, 2001
(1985),\\
 S.J. Brodsky, H.C. Pauli and S.S. Pinsky, Preprint MPIH-V1-1991

\item[2.] T. Maskawa, K. Yamawaki, Progr. Theor. Phys. {\bf 56}, 270 (1976)

\item[3.] R.S. Wittman, in:{\em Nuclear and Particle Physics on the
Light-Cone,}\\
 M.B. Johnson, L.S. Kisslinger, eds., World Scientific,
Singapore 1989

\item[4.] T. Heinzl, S. Krusche, S. Simb\"urger and E. Werner, Z.Phys. {\bf
C56}, 415 (1992)\\
T. Heinzl, C. Stern, E. Werner and B. Zellermann, Z.Phys. {\bf C72},
353 (1996)

\item[5.] G. McCartor and D.G. Robertson, Z.Phys. {\bf C53}, 679 (1992)

\item[6.] D.G. Robertson, Phys.Rev. {\bf D47}, 2549 (1993)

\item[7.] S.S. Pinsky, B. van de Sande and C.M. Bender, Phys.Rev. {\bf D48},
816 (1993)\\
S.S. Pinsky and B. van de Sande,  Phys.Rev. {\bf D49}, 2001 (1994)\\
S.S. Pinsky, B. van de Sande  and J.R. Hiller, Phys.Rev. {\bf D51}, 
726 (1995)\\
A.C. Kalloniatis and H.C. Pauli, Z.Phys. {\bf C63}, 161 (1994)\\
A.C. Kalloniatis and D.G. Robertson,  Phys.Rev. {\bf D50}, 5262 (1994)\\
A.C. Kalloniatis, H.C. Pauli and S.S. Pinsky, Phys.Rev. {\bf D50},
6633 (1994)

 \item[8.] R. Haag, Phys.Rev.{\bf 112}, 669 (1958)

\item[9.] L. Schwartz, Th\'eories des distributions, Hermann, Paris 1966\\
A.M. Gelfand and G.E. Shilov, Generalized functions, Academic Press,
New-York 1967.    

\item[10.] G. Parisi, J.Stat.Phys. {\bf 23}, 49 (1980), Nuovo Cimento {\bf
A21}, 179 (1974)

\item[11.] J. Zinn-Justin, Quantum Field Theory and Critical Phenomena,
Oxford Science Publication, Clarendon Press Oxford 1990, Chap.22.3,
25.1-25.4

\item[12.] F. Cooper et al, Nucl.Phys. {\bf B210} $[FS6]$ 210 (1982)

\item[13.] L. H\"ormander, The Analysis of Linear Partial Differential
Operators I, Springer-Verlag, Berlin 1990

\item[14.] N.N. Bogoliubov, General Principles of Quantum Field Theory,
Kluwer Academic Publishers, Dordrecht 1990.

\end{description}
\end{document}